\documentclass[aps,prl,10pt,twocolumn,superscriptaddress]{revtex4-1}

\usepackage{times}
\usepackage{changes}

\usepackage{graphicx}
\usepackage{amssymb}
\usepackage{epstopdf}
\usepackage{amsmath}
\usepackage{color}
\usepackage{hyperref}
\usepackage{bbold}
\usepackage{mathtools}

\usepackage[acronym]{glossaries}
\usepackage{physics}
\makeatletter

\@ifundefined{textcolor}{}
{
 \definecolor{BLACK}{gray}{0}
 \definecolor{WHITE}{gray}{1}
 \definecolor{RED}{rgb}{1,0,0}
 \definecolor{GREEN}{rgb}{0,1,0}
 \definecolor{BLUE}{rgb}{0,0,1}
 \definecolor{CYAN}{cmyk}{1,0,0,0}
 \definecolor{MAGENTA}{cmyk}{0,1,0,0}
 \definecolor{YELLOW}{cmyk}{0,0,1,0}
}

\graphicspath{{Images/}}

\hypersetup{
	pdfpagemode={UseOutlines},
	bookmarksopen,
	pdfstartview={FitH},
	colorlinks,
	linkcolor={black},
	citecolor={blue},
	urlcolor={blue}}
	
\usepackage{hyperref}
\hypersetup{
colorlinks=true,
linkcolor=black,           
citecolor=blue,           
filecolor=magenta,        
urlcolor=cyan
}

\makeatother

\begin{document}

\title{Higher-order and fractional discrete time crystals in clean long-range interacting systems}

\author{Andrea Pizzi}
\affiliation{Cavendish Laboratory, University of Cambridge, Cambridge CB3 0HE, United Kingdom}
\author{Johannes Knolle}
\email{j.knolle@tum.de}
\affiliation{Department of Physics, Technische Universit{\"a}t M{\"u}nchen, James-Franck-Stra{\ss}e 1, D-85748 Garching, Germany}
\affiliation{Munich Center for Quantum Science and Technology (MCQST), 80799 Munich, Germany}
\affiliation{Blackett Laboratory, Imperial College London, London SW7 2AZ, United Kingdom}
\author{Andreas Nunnenkamp}
\affiliation{Cavendish Laboratory, University of Cambridge, Cambridge CB3 0HE, United Kingdom}

\maketitle

\section{Abstract}
Discrete time crystals are periodically driven systems characterized by a response with periodicity $nT$, with $T$ the period of the drive and $n>1$. Typically, $n$ is an integer and bounded from above by the dimension of the local (or single particle) Hilbert space, the most prominent example being spin-$1/2$ systems with $n$ restricted to $2$. Here we show that a clean spin-$1/2$ system in the presence of long-range interactions and transverse field can sustain a huge variety of different `higher-order' discrete time crystals with integer and, surprisingly, even fractional $n > 2$. We characterize these (arguably prethermal) non-equilibrium phases of matter thoroughly using a combination of exact diagonalization, semiclassical methods, and spin-wave approximations, which enable us to establish their stability in the presence of competing long- and short-range interactions. Remarkably, these phases emerge in a model with continous driving and time-independent interactions, convenient for experimental implementations with ultracold atoms or trapped ions.

\section{Introduction}
Due to its foundational and technological relevance, the study of condensed matter systems out of equilibrium has attracted growing interest in recent years, accounting among others for the discovery of dynamical phase transitions \cite{heyl2013dynamical, heyl2019dynamical}, quantum scars \cite{turner2018weak} and, particularly, discrete time crystals (DTCs) \cite{sacha2015modeling, else2016floquet, von2016phase, von2016absolute, khemani2016phase, sacha2017time, else2019discrete, khemani2019brief, russomanno2017floquet}. A DTC is a non-equilibrium phase of matter breaking the discrete time translational symmetry of a periodic (i.e., Floquet) drive. In the thermodynamic limit, the defining feature of a $n$-DTC is a subharmonic response at $1/n$-th of the drive frequency ($n > 1$), which is robust to perturbations and which persists up to infinite time \cite{russomanno2017floquet}. Following the first seminal proposals \cite{sacha2015modeling, else2016floquet, von2016phase, von2016absolute, khemani2016phase}, DTCs have been widely investigated both theoretically and experimentally \cite{russomanno2017floquet, moessner2017equilibration, else2017prethermal, zhu2019dicke, yao2017discrete, zhang2017observation, choi2017observation, rovny2018observation, smits2018observation, gambetta2019classical, lazarides2019time, gambetta2019discrete}.

In this context, most work has focused on spin-$1/2$ systems, which have largely been shown to exhibit a $2$-DTC where at every Floquet period each spin (approximately) oscillates between the states $\ket{\uparrow}$ and $\ket{\downarrow}$ leading to period doubling ($n=2$). This fact naturally emerges from the dimension $2$ of the local Hilbert space of the spins \cite{von2016phase}, and can be generalized to $n$-DTCs in models of $n$-dimensional clocks \cite{sreejith2016parafermion, surace2018floquet}. Another well-studied setting is that of bosons in a gravitational field bouncing on an oscillating mirror \cite{sacha2017time}, where the single-particle Hilbert space dimension is infinite (as the particle's position is continuous) and where $n$-DTCs with arbitrary integer \cite{giergiel2018time} and fractional \cite{matus2019fractional} $n$ have been shown.

\begin{figure}[t]
	\begin{center}
		\includegraphics[width=\linewidth]{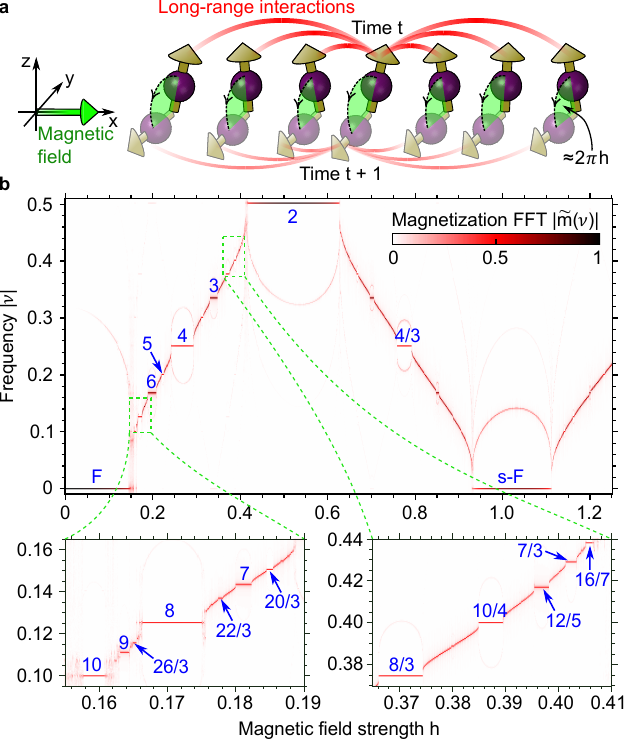}\\
	\end{center}
	\vskip -0.5cm \protect
	\caption{
		\textbf{Higher-order and fractional discrete time crystals.}
		(a) A spin-$1/2$ chain with long-range interactions and initial $\textbf{z}$-polarization is driven with a monochromatic transverse magnetic field of strength $h$, inducing a spin precession around $\textbf{x}$.
		(b) The time crystallinity is probed by the Fourier transform $|\tilde{m}(\nu)|$ of the magnetization along $\textbf{z}$. The spectrum fragments in a multitude of plateaus with constant frequency $1/n$ for a magnetic field strength $h$ in a \textit{finite} range $\approx1/n$, each of which signals a higher-order $n$-DTC robust to perturbations of the drive ($n$ is indicated in blue font for some of the resolved DTCs). Especially remarkable are fractional $n$-DTCs, with $n = p/q$ and $p$ and $q$ some coprime integers. This spectrum refers to the LMG limit ($\alpha = 0, \lambda = 0$), at fixed interaction $J = 0.5$, restricting to the first frequency Brillouin zone $-0.5 \le \nu \le 0.5$, for $500$ and $2000$ drive periods in the top and bottom panels, respectively.
	}
	\label{fig1}
\end{figure}

In these systems, heating to a featureless `infinite temperature' state is typically avoided by introducing disorder, which leads to a (Floquet) many-body-localized (MBL) phase \cite{else2016floquet, khemani2016phase}. Alternatively, in clean (i.e., non-disordered) systems, heating can be escaped with all-to-all interactions \cite{russomanno2017floquet, surace2018floquet, pizzi2019period}, or significantly slowed down with long-range interactions \cite{lerose2019quasilocalized, liu2019confined, tran2019locality, machado2019prethermal}. Very recently, Ref.~\cite{machado2019prethermal} has provided the theoretical framework to study Floquet, clean, long-range interacting systems, in which novel prethermal phases of matter are expected. While their framework allows for the possibility of $n$-DTCs with $n$ larger than the size of the local (or single-particle) Hilbert space, their concrete examples are limited to $n=2$. From our analysis below, we see that part of the difficulty in numerically observing what we call `higher-order' DTCs may lie in their emergence at system sizes which are typically beyond the reach of exact diagonalization.

Here, we overcome this limitation by considering a system amenable to a set of complimentary methods, which enable us to discover an unusually rich dynamical phase diagram hosting a zoo of novel, exotic, (and arguably prethermal) non-equilibrium phases of matter. More specifically, we show that a clean spin-$1/2$ chain in the presence of long-range interactions (Fig.~\ref{fig1}a) can sustain robust higher-order $n$-DTCs with integer and, remarkably, even fractional $n>2$ (e.g., $n = 3, 4, 8/3$ and beyond). These novel dynamical phases give rise to a peculiar fragmentation of the magnetization spectrum, which is intriguingly reminiscent of the plateau structure of the Fractional Quantum Hall Effect.

In the following we present a rather general model of long-range interacting spins, thoroughly study its semiclassical (that is, mean-field) limit, and finally show that the physics observed extends far beyond the fine-tuned limit. We note that our work is distinct from traditional MBL DTCs, which do not have such a semiclassical limit. On a conceptual level our analysis is closer to that of equilibrium statistical physics, where for example the ferromagnetic phase in the Ising model is best understood in a mean-field description which is exact in the limit of all-to-all interactions. In our out-of-equilibrium and clean setting, the existence of a conceptually simple mean-field limit is particularly valuable, and highlights profound differences between the clean DTCs considered here and the pioneering works on MBL DTCs.

\begin{figure*}[t]
	\begin{center}
		\includegraphics[width=\linewidth]{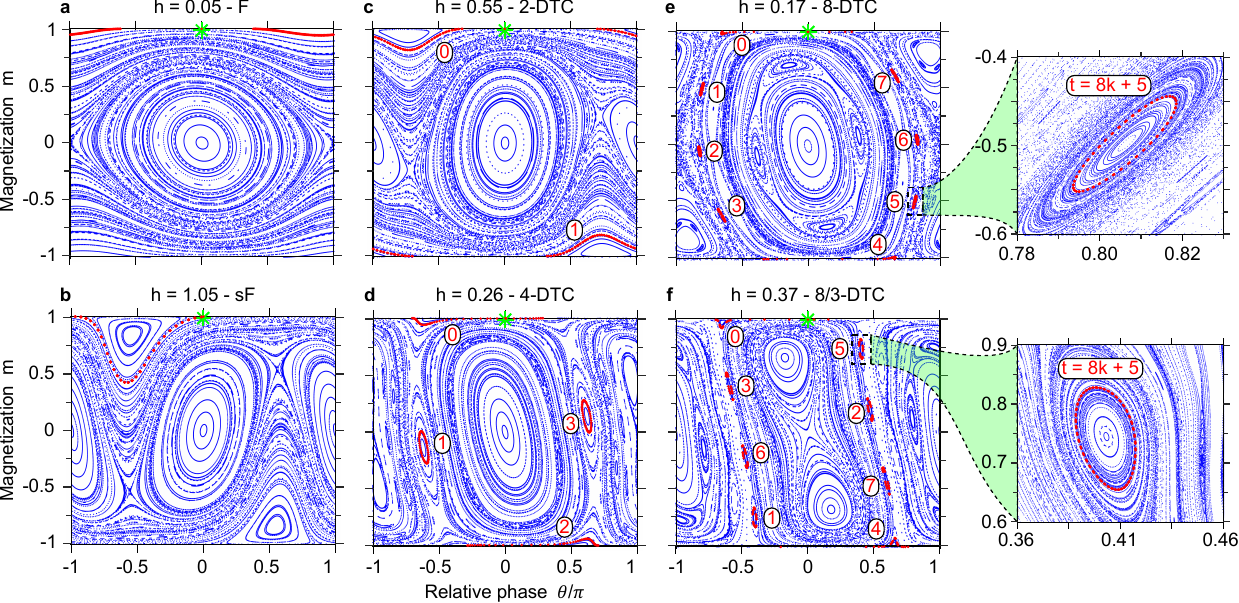}\\
	\end{center}
	\vskip -0.5cm \protect\caption
	{\textbf{Phase space structure of the dynamical phases.} Poincar{\'e} maps of the semiclassical dynamics \eqref{eq. GPE} for various magnetic field strengths $h$ and a fixed interaction $J = 0.5$. Red markers highlight the trajectory starting in the $\textbf{z}$-polarized state ($m = 1, \theta = 0$, green asterisk). (a) dynamical ferromagnet (F): the magnetization $m$ remains $\approx 1$ at all times; (b) stroboscopic ferromagnet (sF): the magnetization $m$ changes sign during the micromotion and yet it remains positive at stroboscopic times; (c) $2$-DTC: the system alternatively visits two islands of the phase space -- one with $m \approx 1$ (numbered as $0$) at even times, and the other with $m \approx -1$ (numbered as $1$) at odd times; (d,e) higher-order $n$-DTCs with integer $n = 4,8$, respectively: the system visits cyclically $n$ islands of the phase space (accordingly numbered in red), with one tour of the islands corresponding to one complete revolution of the spins around the Bloch sphere. (f) Higher-order $n$-DTC with fractional $n = q/p = 8/3$: it takes $p$ revolutions of the spins for the system to tour $q$ islands of the phase space, resulting in a sharp magnetization oscillation frequency $\nu = p/q$. The insets on the right zoom on the island visited at times $t = 8k + 5$, $k = 0, 1, 2, \dots$ for the $8$-DTC (top) and the $8/3$-DTC (bottom).}
	\label{fig2}
\end{figure*}

\section{Results}
We consider a one-dimensional chain of $N$ spins in the thermodynamic limit ($N \rightarrow \infty$), driven according to the following time-periodic Hamiltonian
\begin{equation}
\begin{aligned}
H(t) =
& \frac{J}{\mathcal{N}_{N,\alpha}} \sum_{\substack{i, j = 1 \\ i \neq j}}^{N} \frac{\sigma_i^z\sigma_j^z}{(r_{i,j})^\alpha}
+ \lambda \sum_{j = 1}^{N} \sigma_j^z\sigma_{j+1}^z \\
& - \pi h[1+\sin(2\pi t)] \sum_{j = 1}^{N} \sigma_j^x,
\end{aligned}
\label{eq. H}
\end{equation}
where $\sigma_j^{(x,y,z)}$ denote the standard Pauli operators for the $j$-th spin, periodic boundary conditions are assumed, and both $\hbar$ and the drive frequency have been set to $1$. $J$ measures the strength of a power-law interaction with characteristic exponent $\alpha$, $\lambda$ is the strength of a nearest-neighbor interaction, and $\pi h$ is the average over one drive period of the monochromatic transverse magnetic field. The Kac normalization $\mathcal{N}_{N, \alpha} = \sum_{j=2}^{N} \frac{1}{(r_{1,j})^\alpha}$ guarantees extensivity, and conveniently allows to stretch the model to the Lipkin-Meshkov-Glick (LMG) limit of all-to-all interactions ($\alpha = \lambda = 0$), in which the underlying complex physics is reduced to its essence and most easily interpreted.

The dynamics from an initially $\textbf{z}$-polarized state $\ket{\psi(0)} = \ket{\uparrow, \uparrow, \dots, \uparrow}$ is integrable in the non-interacting limit $J = \lambda = 0$, for which the magnetization $m(t) = \langle \sigma_j^z \rangle(t)$ at stroboscopic times $t = 0,1,2,\dots$ reads $m(t) = \cos(2\pi ht)$, that is $h$ is the system's characteristic frequency. The essential question to diagnose a $n$-DTC is whether, upon switching on the interactions, there exists a finite range of $h$ for which the system's characteristic frequency $\nu$ remains instead locked to a constant value $1/n < 1$, signalling the stability of the DTC to perturbations of the drive. In the following we answer this question affirmatively not only for the well-known $n=2$ case, but, if the interactions are sufficiently long-range, also for integer and even fractional $n>2$, corresponding to the higher-order DTCs. Note, the $\mathbb{Z}_2$ symmetry of the Hamiltonian in Eq.~\eqref{eq. H} raises, for a $2$-DTC, the conceptual issue whether the subharmonic response stems from the time-symmetry breaking itself or it rather ``piggybacks'' on an underlying breaking of the $\mathbb{Z}_2$ symmetry \cite{von2016absolute, else2017prethermal, else2020long}. This issue however disappears for the higher-order $n$-DTCs that, because the Hamiltoanian lacks any $\mathbb{Z}_n$ symmetry ($n>2$), must indeed be a ``genuine'' manifestation of time-symmetry breaking.

For the sake of clarity, we first focus on the LMG limit of all-to-all interactions ($\alpha = \lambda = 0$), which allows for a conceptually simple semiclassical interpretation of the various dynamical phases. The dynamics of the system is in this case described by a semiclassical Gross-Pitaevskii equation (GPE) for the complex fields $\psi_\uparrow$ and $\psi_\downarrow$ (details in Supplementary Note 1)
\begin{equation}
\begin{aligned}
\frac{d \psi_\uparrow}{d(it)} &= \pi h[1 + \sin2\pi t] \psi_\downarrow - 4J |\psi_\uparrow|^2\psi_\uparrow, \\
\frac{d \psi_\downarrow}{d(it)} &= \pi h[1 + \sin2\pi t] \psi_\uparrow - 4J |\psi_\downarrow|^2\psi_\downarrow,
\end{aligned}
\label{eq. GPE}
\end{equation}
where we can identify $|\psi_\uparrow|^2-|\psi_\downarrow|^2 \rightarrow m = \langle \sigma_j^z \rangle $ and $\psi_\uparrow^* \psi_\downarrow = |\psi_\uparrow||\psi_\downarrow| e^{i\theta} \rightarrow \frac{\langle \sigma_j^x \rangle + i \langle \sigma_j^y \rangle}{2}$. While in the limit $\alpha \to 0$ the dynamics described by the GPE \eqref{eq. GPE} is indeed $0$-dimensional and lacks any sense of locality, it carries the signature of many-body interactions in its nonlinearity rather than in an exponentially large number of degrees of freedom (similarly, e.g., to the paradigmatic mean-field equation $m = \tanh[Jm/k_B T]$ of the Ising model in equilibrium). Note, the presence of such a limit highlights qualitative differences between clean long-range DTCs and MBL DTCs, and is at the heart of their much richer phenomenology.

The dynamics of the magnetization $m$ is obtained integrating the GPE \eqref{eq. GPE} from an initially $\textbf{z}$-polarized state ($\psi_\uparrow(0) = 1, \psi_\downarrow(0) = 0$), and the corresponding Fourier transform $|\tilde{m}(\nu)|$ versus the magnetic field strength $h$ is plotted in Fig.~\ref{fig1}b. As it is well-known \cite{russomanno2017floquet}, the $2$-DTC results in the system characteristic frequency $\nu$ being locked to $1/2$ for $h \approx 1/2$. Surprisingly, the same locking occurs at frequencies $1/n$ with integer and fractional $n>2$ (e.g.~$n = 3, 4, 8/3$), giving rise to a fragmentation of the spectral line of $\tilde{m}(\nu)$ in a sequence of plateaus of constant frequency for a finite range of $h \approx 1/n$. Each of these plateaus signals a higher-order (possibly fractional) DTC, the width of the plateau being a signature of the DTC's robustness to drive perturbations. The `halos' surrounding the plateaus in Fig.~\ref{fig1} correspond to incommensurate (non-subharmonic) frequencies adding a time-glassy aspect to the DTCs. The magnitude of these secondary peaks is in the order of a few percent compared to the dominant subharmonic peak, resulting in weak aperiodic fine features on top of the subharmonic response.

The plateau at $\nu = 0$ for $h \approx 0$ signals the tendency of the spins to remain aligned along $\textbf{z}$ in a dynamical ferromagnetic phase (F). This corresponds to macroscopic quantum self-trapping of weakly driven bosons in a double well \cite{albiez2005direct}, which can in fact be exactly mapped to the LMG limit (details in Supplementary Note 1). For $h \approx 1,2,3,\dots$, the spins complete approximately $1,2,3,\dots$ revolutions around the Bloch sphere at each drive period, respectively, and yet maintain a preferential alignement along $\textbf{z}$ at stroboscopic times, in what may be called a stroboscopic-ferromagnetic phase (sF).

Our results are confirmed by exact diagonalization studies. Thanks to the all-to-all coupling of the LMG limit, the dynamics is in fact confined to the symmetric sector, whose size grows only linearly with the number of spins $N$. This allows a scaling analysis extended up to large system sizes, showing a progressive emergence of the spectral line plateaus for an increasing number of spins $N$. For the standard $2$-DTC, the plateau is clearly visible already for $N \gtrapprox 10$, whereas, crucially, for the $4$-DTC it appears only for $N \gtrapprox 100$ (see details in Supplementary Note 2). This observation strongly suggests that signatures of the higher-order $n$-DTCs arise for larger system sizes as compared to the standard $2$-DTC, making them generally elusive to exact diagonalization techniques. This fact might explain the difficulties in observing higher-order DTCs in the past and motivates the choice of model \eqref{eq. H} in the first place.

The stroboscopic dynamics generated by the GPE \eqref{eq. GPE} can be conveniently described with Poincar{\'e} maps, popular tools in dynamical systems theory that here provide an immediate interpretation of much of the underlying physics, which to some extent characterizes the dynamical phases also when deviating from the LMG limit. In Fig.~\ref{fig2}, the trajectory starting in the $\textbf{z}$-polarized state (green asterisk) is highlighted with red markers. For a weak drive $h \approx 0$, the spins tend to remain aligned along $\textbf{z}$ in a dynamical ferromagnetic phase (a), giving rise to a Poincar{\'e} map which closely resembles the phase portrait of undriven bosons in a double well \cite{albiez2005direct}. For $h \approx 1$, the micromotion consists of approximately an entire revolution of the spins around the Bloch sphere per period, with a preferential $\textbf{z}$ alignment restored at stroboscopic times despite the detuning in the magnetic field strength (b). For $h \approx 1/n$ and $n = 2,4,8$ in (c), (d) and (e), respectively, the $n$-DTC results in the presence of $n$ `islands' in the phase space which the system visits sequentially jumping from one to the next at each drive period. In $n$ drive periods, the system visits all the $n$ islands once, and the magnetization $m$ completes one oscillation.

Furthermore, for $h \approx 3/8$ the system behaves as a $n$-DTC with fractional $n = q/p = 8/3$ (f). In this case, the system cyclically visits $q$ islands of the phase space in $q$ drive periods. Differently from a $q$-DTC, however, during this time the magnetization $m$ completes $p$ oscillations, resulting in a characteristic frequency $p/q$. Finally, for larger interactions $J$ the Poincar{\'e} maps become chaotic (as in Ref.~\cite{russomanno2017floquet}), signaling thermalization \cite{cosme2014thermalization}.

Since the island-to-island hopping that underpins the subharmonic response holds for any point of any island, the islands themselves can be interpreted as stability regions of the DTCs with respect to perturbations of the initial state. The Poincar{\'e} maps also provide an interpretation of the subdominant frequencies visible as halos around the plateaus in Fig.~\ref{fig1}, that are in fact associated with the revolution period of the intra-island orbits (e.g., those shown in the insets of Fig.~\ref{fig2}). This also explains why these frequencies are sensitive to perturbations of the drive, which deform the shape of the islands and thus the orbits’ revolution periods, but are only weakly sensitive to perturbations of the initial state.

\begin{figure}[t]
	\begin{center}
		\includegraphics[width=1\linewidth]{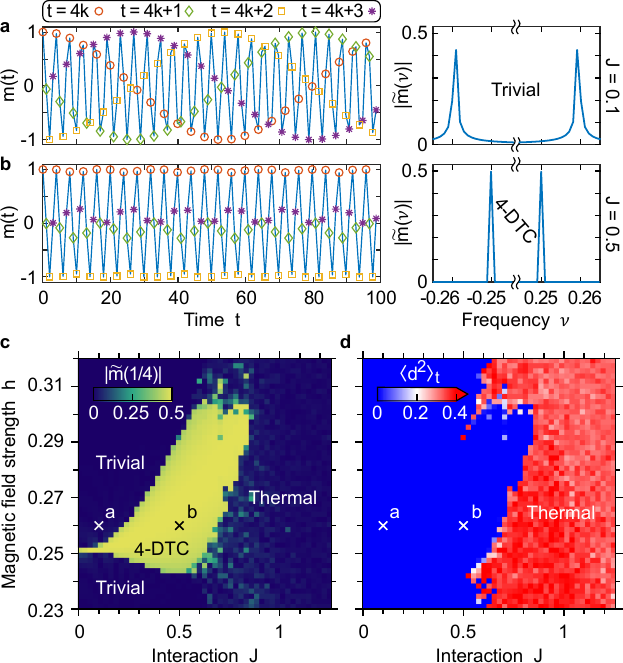}\\
	\end{center}
	\vskip -0.5cm \protect\caption
	{\textbf{Many-body nature of the higher-order discrete time crystals.} The robustness of the higher-order time crystals is induced by the interactions, justifing their classification as non-equilibrium phases of matter. For concreteness, we show this for the $4$-DTC in the LMG limit. (a,b) Magnetization $m(t)$ at stroboscopic times (left) and respective Fourier transform $|\tilde{m}(\nu)|$ (right) for a slightly detuned magnetic field strength $h = 1/4 + 0.01$. For a weak interaction $J = 0.1$ in (a), the system trivially oscillates at frequencies $\nu \approx \pm 0.26$, whereas a larger interaction $J = 0.5$ in (b) re-establishes a robust subharmonic response at frequency $\nu = 1/4$. (c) Amplitude of the subharmonic peak $|\tilde{m}(1/4)|$ in the $(J,h)$ plane. The $4$-DTC phase opens up from the integrable point $J = 0, h = 0.25$, that is the interaction makes the $4$-DTC robust. (d) A thermal region of the phase space is characterized by a finite value $\sim 1$ of the decorrelator time-average $\langle d^2 \rangle_t$, corresponding to semiclassical chaos. Both (c) and (d) are computed over $10^3$ drive periods.}
	\label{fig3}
\end{figure}

It is well-established for the standard $2$-DTC that the robust subharmonic response hinges on the interaction being sufficiently strong. The fact that interactions are necessary for the robustness of DTCs is critical, as it underpins the many-body nature of the DTCs and it justifies their classification as non-equilibrium phases of matter \cite{else2019discrete}. It becomes thus of primary importance to assess the role of the interactions also for the higher-order DTCs. To this end, as a concrete example, in Fig.~\ref{fig3} we investigate the effects of the interaction strength $J$ on the $4$-DTC. If the interaction is weak, a slightly mistaken magnetic field strength $h = 1/4 + \epsilon$, with $\epsilon \ll 1$, originates in envelopes (that is, beatings) with period $\sim 1/\epsilon$, resulting in the Fourier transform $\tilde{m}$ being peaked at $\nu \approx h$ and in trivial dynamics (a). Crucially, stronger interactions can compensate the mistake in the flipping field (b): the envelopes in $m(t)$ disappear, the peak in $\tilde{m}$ is set back to the subharmonic frequency $\nu = 1/n$, and the discrete time symmetry is broken. The time-glassy character of the DTC is observed in a small aperiodic modulation of the magnetization on top of the subharmonic response.

The subharmonic peak magnitude $|\tilde{m}(1/4)|$ can be used to trace out the $4$-DTC phase in the ($J,h$) plane (c). The $4$-DTC phase opens up from the integrable point $J = 0, h = 1/4$ for increasing interactions, in analogy with the opening of the standard $2$-DTC from $J = 0, h = 1/2$ \cite{yao2017discrete}. This opening, which in dynamical system theory would be referred to as Arnold's tongue, confirms that larger interactions $J$ allow the higher-order DTCs to bear larger detunings in the field $h$. However, at even larger $J \gtrapprox 0.8$ semiclassical chaos sets in and the time crystalline order is broken irrespectively of $h$. To see this, we introduce a decorrelator $\langle d^2(t) \rangle_t$ (see Methods and Supplementary Note 3), measuring the average distance between two initially very close copies of the system evolving under Eq.~\eqref{eq. GPE}. $\langle d^2 \rangle_t \sim 1$ corresponds to sensitivity to the initial conditions, that is, to classical chaos, which in turn signals quantum thermalization \cite{cosme2014thermalization}.

As shown, the DTCs rely on the interactions being sufficiently (but not too) strong. Crucially, in contrast to the standard $2$-DTC, higher-order DTCs also necessitate the interactions to be sufficiently long-range. We now probe the robustness of the higher-order DTCs along yet a different direction in the drive space, exploring the effects of non-all-to-all interaction on higher-order DTCs, particularly assessing their stability upon breaking the mean-field solvability of the dynamics with power-law ($\alpha>0$) and nearest-neighbor ($\lambda > 0$) interactions. In this case, the system is no longer described as a collective spin, and spin-wave excitations are rather generated. To account for them, we adopt a spin-wave approximation (see Methods), in which the central dynamical variable is the density of spin-wave excitations $\epsilon(t)$
\begin{equation}
\epsilon = \frac{2}{N} \sum_{k \neq 0}^{N} \langle b_k^\dagger b_k \rangle,
\end{equation}
where $b_k^\dagger$ and $b_k$ are bosonic creation and annihilation operators for the spin-waves excitations with momentum $k$.

\begin{figure}[t]
	\begin{center}
		\includegraphics[width=1\linewidth]{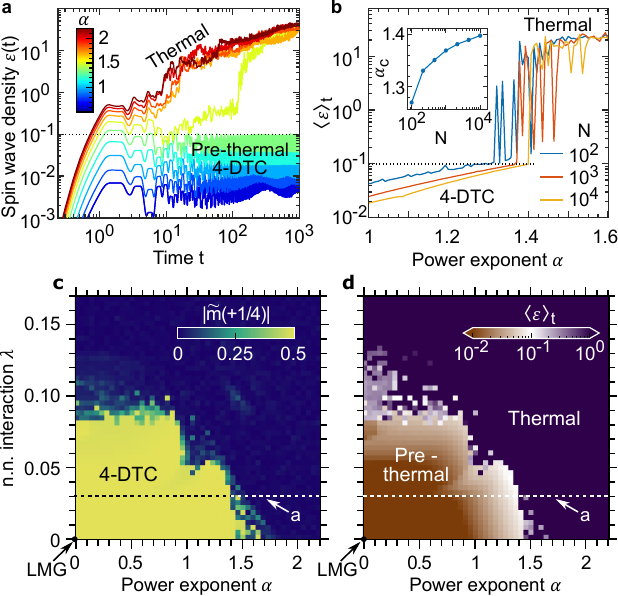}\\
	\end{center}
	\vskip -0.5cm \protect\caption
	{\textbf{Stability and prethermalization with power-law and nearest-neighbor interactions}. The higher-order and fractional DTCs survive, most likely in a prethermal fashion, when deviating from the LMG limit. For concreteness, we focus on the $4$-DTC at $h = 0.27$ and $J = 0.5$, and consider the effects of power-law ($\alpha > 0$) and nearest-neighbor ($\lambda > 0$) interactions. (a) If the interactions are sufficiently long-range (that is $\alpha$ is small enough, here for a fixed $\lambda = 0.03$), the density of spin-wave excitations $\epsilon$ remains small throughout several time decades. Conversely, shorter-range interactions lead to the proliferation of spin-wave excitations that makes the system quickly thermalize destroying any time crystalline order \cite{zhu2019dicke}. (b) A sharp transition between these two regimes is highlighted by the time-average $\langle \epsilon \rangle_t$ over $10^3$ periods versus $\alpha$ (at a fixed $\lambda = 0.03$). The critical $\alpha_c$ at which $\langle \epsilon \rangle_t$ crosses the threshold $0.1$ (inset), grows and possibly saturates with the system size $N$, suggesting the stability of the $4$-DTC in the thermodynamic limit $N\rightarrow\infty$. (c,d) The stability of the $4$-DTC for a whole region of the parameter space surrounding the LMG point $\alpha = \lambda = 0$ is highlighted plotting the magnitude of the subharmonic peak $|\tilde{m}(1/4)|$ and the average spin-wave density $\langle \epsilon \rangle_t$ in the $(\alpha, \lambda)$ plane.}
	\label{fig4}
\end{figure}

In the LMG limit ($\lambda = \alpha = 0$), no spin-wave excitation is generated and $\epsilon = 0$ at all times. When departing from such a limit, two scenarios are possible (Fig.~\ref{fig4}a): (i) $\epsilon$ rapidly reaches a plateau $\lessapprox 0.1$ (up to some small fluctuations), for which we consider the spin-wave approximation consistent, or (ii) $\epsilon$ rapidly grows to values $\gtrapprox 1$, for which the spin-wave approximation breaks down. Although the method is not exact and may fail to capture the very long-time physics, it suggests that (i) and (ii) correspond to prethermalization and thermalization, respectively \cite{lerose2018chaotic, lerose2019impact, zhu2019dicke}.

We observe that the higher-order DTCs are stable (at least in a prethermal fashion) for sufficiently long-range interactions (that is, sufficiently small $\lambda$ and $\alpha$), whereas thermalization quickly sets in for shorter-range interactions (Fig.~\ref{fig4}a). The transition between these two dynamical phases is sharp and can be located comparing the spin-wave density time average $\langle \epsilon \rangle_t$ with a threshold $0.1$ (Fig.~\ref{fig4}b). The stability of the $n$-DTC in the presence of competing power-law and nearest-neighbor interactions can be investigated in the $(\alpha, \lambda)$ plane plotting the amplitude of the subharmonic peak $|\tilde{m}(1/n)|$ in Fig.~\ref{fig4}c and the time-averaged spin-wave density $\langle \epsilon \rangle_t$ in Fig.~\ref{fig4}d. The $n$-DTC is stable for a whole region of the parameter space surrounding the LMG point ($\alpha = \lambda = 0$), that is, if the interactions are sufficiently long-range. The DTC is also robust to arbitrary perturbations to the initial state, as long as the initial spin-wave density $\epsilon(0)$ is sufficiently small, as we have checked by injecting a small amount of spin-wave excitations to the initial state. For completeness, in the Supplementary Note 5 we also investigate the stability of the $4$-DTC against the introduction of a longitudinal field of strength $h_z$. We find in this case that the expected prethermal nature of the DTC is only changed quantitatively, with the heating time scaling exponentially with the ratio of the driving frequency (for us always $1$) and the energy scale $h_z$. Finally, we note that the time-glassy character of the DTCs is maintained throughout their whole stability region. In particular, we find small peaks at incommensurate frequencies that do (do not) vary when slightly perturbing the drive (initial conditions).

\section{Discussion}
Higher-order DTCs in clean long-range interacting systems are qualitatively distinct from DTCs of MBL Floquet systems \cite{machado2019prethermal}. Indeed, the higher-order DTCs require the establishment of order along directions different from $\pm \textbf{z}$. For instance, in the $4$-DTC the spins are approximately aligned along $-\textbf{y}$ and $+ \textbf{y}$ at times $t = 1,5,\dots$ and $t = 3,7,\dots$, respectively. In an MBL system, a disordered magnetic field or short-range interaction along $\textbf{z}$ would immediately scramble the system when the spins are far from the $\textbf{z}$ axis, precluding the possibility of higher-order DTCs. Thus, our work establishes that translationally-invariant systems with long-range interactions can circumvent these limitations \cite{else2017prethermal, machado2019prethermal}.

In the LMG limit of all-to-all interactions, the model \eqref{eq. H} has a low-dimensional semiclassical limit, which links the $n$-DTCs to the multifrequency mode locking of some nonlinear discrete maps, which is ubiquitous in the natural sciences \cite{jensen1983complete, belykh1977shunted, guevara1982phase, bramble1983running}. On the one hand, our work establishes a connection between this class of DCTs and dynamical system theory. On the other hand, it provide evidence for the stability of the higher-order DTCs in a whole region of the parameter space surrounding the LMG limit, in the presence of competing, mean-field breaking, long- and short-range interactions, that is in a genuinely quantum setting with no semiclassical counterpart.

The choice of a continuous Floquet drive with constant-in-time interactions and monochromatic transverse magnetic field, together with the translational invariance, makes model \eqref{eq. H} a prime candidate for experimental implementation. For instance, bosons in a double well \cite{albiez2005direct} could be used to realize a truly all-to-all interacting, that is the LMG, model. In this case, the field pulses would be simply implemented lowering the barrier between the two wells to allow particle tunnelling, and the fact that no time modulation for the particle-particle interaction is necessary should provide a major simplification. Power-law interactions with tunable alpha $0 \le \alpha \le 3$ can instead be realized in trapped-ion experiments \cite{islam2013emergence, britton2012engineered,zhang2017observation}. For completeness we note, anyway, that a phenomenology similar to that presented here also emerges in the case of a binary drive (see Supplementary Note 4 for details).

In conclusion, we have discovered higher-order DTCs with a period that is not limited from above by the size of the local (or single-particle) Hilbert space. The dynamical phase space fragments to host many higher-order $n$-DTCs with integer and even fractional $n$, at least in a prethermal fashion. Future work should attempt to gain further analytical understanding regarding the role of long-range interactions in stabilizing the different higher-order DTCs. Most importantly, it should be assessed what are the allowed fractions $q/p$ that result in a $q/p$-DTC?  Further study should assess in more detail the role of the Kac normalization and of the dimensionality on the fate of the dynamical phases of matter presented here.

\section{Methods}

\textbf{Decorrelator.}
We quantify the sensitivity to the initial conditions of Eq.~\eqref{eq. GPE} with a decorrelator $d^2(t)$ \cite{bilitewski2018temperature, pizzi2019period}
\begin{equation}
d^2(t) = \left[|\psi_\uparrow(t)|^2 - |\psi'_\uparrow(t)|^2\right]^2 + \left[|\psi_\downarrow(t)|^2 - |\psi'_\downarrow(t)|^2\right]^2,
\end{equation}
measuring the distance in time between two initially very close copies of the system evolving under Eq.~\eqref{eq. GPE}. Specifically, we consider the following two close initial conditions
\begin{align}
\psi_\uparrow(0) = 1&,
\quad
\psi_\downarrow(0) = 0, \\
\psi_\uparrow'(0) = \cos(\Delta_m) e^{-i\frac{\Delta_\theta}{2}}&,
\quad
\psi_\downarrow'(0) = \sin(\Delta_m) e^{+i\frac{\Delta_\theta}{2}},
\end{align}
with $\Delta_m = \Delta_\theta = 10^{-6}$. The decorrelator time-average $\langle d^2 \rangle_t$ is then given by
\begin{equation}
\langle d^2 \rangle_t = \frac{1}{T+1} \sum_{t = 0}^{T} d^2(t),
\end{equation}
where $T$ is the total simulation time, e.g. $T = 10^3$ in Fig.~\ref{fig3} and Fig.~\ref{fig4}.

\textbf{Spin-wave approximation.}
Here, we briefly summarize the idea behind the spin-wave approximation, which is thoroughly explained in Refs.~\cite{lerose2018chaotic, lerose2019impact} and the supplementary material therein, to which we redirect the reader for further details. In a DTC evolving from an initially $\textbf{z}$-polarized state $\ket{\psi(0)} = \ket{\uparrow, \uparrow, \dots, \uparrow}$, the spins are mostly aligned at all times. Imperfections in the alignment can be described within a Holstein-Primakoff transformation in terms of bosonic spin-wave quasiparticles $b_k^\dagger$. Crucially, the collective spin $\vec{S} = \frac{1}{N} \sum_{j=1}^{N} \vec{\sigma}_j$ rotates in time, so that the Holstein-Primakoff transformation has to be performed in a rotating frame $\mathcal{R}' = (X, Y, Z)$ such that the, say, $Z$ axis tracks the orientation of the collective spin at all times. This tracking is encoded in the condition $\langle S^X \rangle = \langle S^Y \rangle = 0$, from which the dynamics of the rotating frame is obtained self-consistently. The Holstein-Primakoff transformation from spin degrees of freedom to bosonic degrees of freedom is then performed as $\sigma_j^X \rightarrow b_j + b_j^\dagger$, $\sigma_j^Y \rightarrow -i(b_j - b_j^\dagger)$, and the spin-waves degrees of freedom are obtained after a Fourier transform $\tilde{b}_k = \frac{1}{\sqrt{N}} \sum_{j = 1}^{N}e^{-ikj}b_j$. On top of this, an approximation is made in that the Hamiltonian is expanded to lowest non-trivial order in the density of spin-wave excitations $\epsilon = \frac{2}{N} \sum_{k \neq 0}^{N} \langle b_k^\dagger b_k \rangle$, which should remain $\ll 1$ for the approximation to be consistent. This procedure results in a set of $2N$ ordinary differential equations similar to Eqs.~(26) and (29) in the Supplemental Material of Ref.~\cite{lerose2018chaotic}, describing the rotation of the new reference frame and the dynamics of the spin waves at the various momenta.

Finally, we remark the main differences between the implementation of the spin-wave approximation in our work and in Ref.~\cite{lerose2018chaotic}. First, Ref.~\cite{lerose2018chaotic} considers a constant-in-time Hamiltonian, whereas the parameters of the Hamiltonian in the present work are time-dependent. As a consequence, the parameters in the system of ordinary differential equations become time-dependent. Second, Ref.~\cite{lerose2018chaotic} considers a nearest-neighbor interaction on top of an all-to-all one, whereas we consider the more general case of nearest-neighbor interaction on top of a power-law one. The $\cos k$ that appears in the equations of Ref.~\cite{lerose2018chaotic} is therefore substituted by a more generic $\tilde{\mathcal{J}}_k = \sum_{j = 1}^{N} \mathcal{J}_{r_{1,j}} e^{-i r_{1,j} k}$ in ours, where $\mathcal{J}_{r_{i,j}}$ contains both the nearest-neighbor and the power-law interactions.

\textbf{Data availability.}
No datasets were generated or analysed during the current study.

\textbf{Acknowledgements.}
It is a pleasure to thank F.~Carollo, E.~I.~R. Chiacchio, F.~M.~Gambetta, J.~P.~Garrahan and A.~Lazarides for useful discussions. A.~P.~acknowledges support from the Royal Society. A.~N.~holds a University Research Fellowship from the Royal Society and acknowledges additional support from the Winton Programme for the Physics of Sustainability.

\textbf{Author contributions}
A.P. carried out the research, J.K. and A.N. supervised it. All authors gave critical contributions to the manuscript preparation.

\textbf{Data availability.}
No datasets were generated or analysed during the current study.

\textbf{Code availability.}
The codes that support the findings of this study are available from the authors on reasonable request.

\bibliography{longrange_HODTC_biblio}
\bibliographystyle{naturemag}

\clearpage
\newpage
\thispagestyle{empty}

\setcounter{equation}{0}
\setcounter{figure}{0}
\setcounter{page}{1}
\makeatletter 
\renewcommand{\thefigure}{S\arabic{figure}}
\renewcommand{\theequation}{S\arabic{equation}}
\setlength\parindent{10pt}

\onecolumngrid

\begin{center}
	{\fontsize{12}{12}\selectfont
		\textbf{Supplementary Information for "Higher-order and fractional discrete time crystals in clean long-range interacting systems"\\[5mm]}}
	{\normalsize Andrea Pizzi, Johannes Knolle and Andreas Nunnenkamp \\[1mm]}
\end{center}
\normalsize

	The Supplementary Information are devoted to technical details of the derivations and complimentary results and is organized as follows. In Supplementary Note 1 we show the mapping of the Lipkin-Meshkov-Glick (LMG) model for fully-connected spins to a model of bosons in a double well, and exploit it to obtain the Gross-Pitaevskii equation (GPE) in the limit of inifinite number of spins $N$. In Supplementary Note 2 we present results from exact diagonalization, showing how the discrete time crystals (DTCs) emerge for an increasing number of spins $N$. We show that higher-order $n$-DTCs emerge at much larger $N$ as compared to the standard $2$-DTC, which explains the difficulty to numerically observe them. In Supplementary Note 3 we give an overview of the spin-wave approximation we employ, whereas in Supplementary Note 4 we show that a binary Floquet protocol also leads to results similar to the ones of the continuous driving discussed in the main text. In Supplementary Note 5 we present an analysis on the stability of the higher-order $n$-DTCs in the presence of a longitudinal field breaking the $\mathbb{Z}_2$ symmetry of the Hamiltonian, showing that the absolute stability persists in the LMG limit and the expected prethermal nature of the DTC beyond this limit is only changed quantitatively.

\clearpage

\section*{Supplementary Note 1: GROSS-PITAEVSKII EQUATION}
In this Note, we derive the semiclassical GPE of motion for the LMG model of $N$ fully-connected spins ($\lambda = \alpha = 0$). To this end, we first map the spin system into a model of bosons in a double well. We then derive Heisenberg equations for the bosonic operators, and we treat them semiclassically replacing the bosonic operators with complex numbers.

\subsection{Map to bosons in a double well}
The Schwinger's oscillator model of angular momentum connects the algebra of angular momentum and the algebra of two bosonic modes \cite{sakurai1995modern}. Here, we show this connection explicitly for the collective spin of a system of $N$ spin-$1/2$.

Given $p = (p_1, p_2, \dots, p_N)$ a permutation of the indexes $1,2,\dots, N$, we say $P_p$ the permutation operator acting as
\begin{equation}
P_p |s_1, s_2, \dots, s_N\rangle = |s_{p_1}, s_{p_2}, \dots s_{p_N}\rangle,
\end{equation}
where $s_i \in \{\uparrow, \downarrow\}$. The permutation operators are used to build the symmetrization operator $\mathcal{S}$, defined as
\begin{equation}
\mathcal{S} = \frac{1}{\sqrt{N!}}\sum_{p} P_p.
\end{equation}
We say $\ket{n_\uparrow, n_\downarrow}$ the symmetrized state with $n_\uparrow$ spins up and $n_\downarrow$ spins down, that is
\begin{equation}
\ket{n_\uparrow, n_\downarrow} = \frac{1}{\sqrt{n_\uparrow!n_\downarrow!}} \mathcal{S} |\underbrace{\uparrow, \uparrow, \dots, \uparrow}_{n_\uparrow}, \underbrace{\downarrow, \downarrow, \dots, \downarrow}_{n_\downarrow}\rangle.
\end{equation}
Since $N$ is fixed and $n_\downarrow = N - n_\uparrow$, the notation $\ket{n_\uparrow, n_\downarrow}$ is actually slightly redundant. In the following, we nevertheless prefer to stick with this notation for the sake of clarity. The states $\ket{n_\uparrow, n_\downarrow}$ form a basis for the Hilbert subspace of completely symmetrized states. Given an operator $O$ commuting with the symmetrization operator $\mathcal{S}$, the action of $O$ on this subspace is therefore fully-characterized by its action on the states $\ket{n_\uparrow, n_\downarrow}$.

In particular, we consider the 'collective' operators $\left(\sum_{j = 1}^{N} \sigma_j^\alpha\right)$ with $\alpha = x,y,z$, which indeed all commute with $\mathcal{S}$.

We have
\begin{align}
\left(\sum_{j = 1}^{N} \sigma_j^x\right) \ket{n_\uparrow, n_\downarrow} = & \frac{1}{\sqrt{n_\uparrow!n_\downarrow!}} \mathcal{S} \sum_{j = 1}^{N} \sigma_j^x |\underbrace{\uparrow, \dots, \uparrow}_{n_\uparrow}, \underbrace{\downarrow, \dots, \downarrow}_{n_\downarrow}\rangle\\
= &\frac{1}{\sqrt{n_\uparrow!n_\downarrow!}}
\sum_{j = 1}^{n_\uparrow} \mathcal{S} |\underbrace{\uparrow, \dots, \uparrow}_{n_\uparrow-1}, \underbrace{\downarrow, \dots, \downarrow}_{n_\downarrow+1}\rangle +
\frac{1}{\sqrt{n_\uparrow!n_\downarrow!}}\sum_{j = n_\uparrow + 1}^{N} \mathcal{S} |\underbrace{\uparrow, \dots, \uparrow}_{n_\uparrow+1}, \underbrace{\downarrow, \dots, \downarrow}_{n_\downarrow-1}\rangle\\
=&n_\uparrow\frac{\sqrt{(n_\uparrow-1)!(n_\downarrow+1)!}}{\sqrt{n_\uparrow!n_\downarrow!}} \ket{n_\uparrow - 1, n_\downarrow + 1} +
n_\downarrow\frac{\sqrt{(n_\uparrow+1)!(n_\downarrow-1)!}}{\sqrt{n_\uparrow!n_\downarrow!}} \ket{n_\uparrow + 1, n_\downarrow - 1}\\
= & \sqrt{n_\uparrow(n_\downarrow + 1)} \ket{n_\uparrow - 1, n_\downarrow + 1} + \sqrt{(n_\uparrow + 1)n_\downarrow} \ket{n_\uparrow + 1, n_\downarrow - 1},
\end{align}
and, similarly,
\begin{align}
\left(\sum_{j = 1}^{N} \sigma_j^y\right) \ket{n_\uparrow, n_\downarrow}
= &\frac{i}{\sqrt{n_\uparrow!n_\downarrow!}}
\sum_{j = 1}^{n_\uparrow} \mathcal{S} |\underbrace{\uparrow, \dots, \uparrow}_{n_\uparrow-1}, \underbrace{\downarrow, \dots, \downarrow}_{n_\downarrow+1}\rangle -
\frac{i}{\sqrt{n_\uparrow!n_\downarrow!}}\sum_{j = n_\uparrow + 1}^{N} \mathcal{S} |\underbrace{\uparrow, \dots, \uparrow}_{n_\uparrow+1}, \underbrace{\downarrow, \dots, \downarrow}_{n_\downarrow-1}\rangle\\
= & i \sqrt{n_\uparrow(n_\downarrow + 1)} \ket{n_\uparrow - 1, n_\downarrow + 1} - i \sqrt{(n_\uparrow + 1)n_\downarrow} \ket{n_\uparrow + 1, n_\downarrow - 1}
\end{align}
and, finally,
\begin{equation}
\left(\sum_{j = 1}^{N} \sigma_j^z\right) \ket{n_\uparrow, n_\downarrow} =
(n_\uparrow - n_\downarrow) \ket{n_\uparrow, n_\downarrow}.
\end{equation}
Introducing standard bosonic operators $a_\uparrow, a_\downarrow, a_\uparrow^\dagger, a_\downarrow^\dagger$ for the two bosonic modes labeled by $\uparrow$ and $\downarrow$, we can thus write
\begin{equation}
\sum_{j = 1}^{N} \sigma_j^x = a_\uparrow^\dagger a_\downarrow + a_\downarrow^\dagger a_\uparrow,
\qquad
\sum_{j = 1}^{N} \sigma_j^y = -i\left(a_\uparrow^\dagger a_\downarrow - a_\downarrow^\dagger a_\uparrow\right),
\qquad
\sum_{j=1}^{N} \sigma_j^z = n_\uparrow - n_\downarrow
\label{eq. map}
\end{equation}
with $n_\uparrow = a_\uparrow^\dagger a_\uparrow$ and $n_\downarrow = a_\downarrow^\dagger a_\downarrow$.

The Hamiltonian (1) in the LMG limit ($\alpha = \lambda = 0$) reads
\begin{equation}
H = \frac{J}{N} \sum_{i,j}^{N} \sigma_i^z\sigma_j^z - \pi h [1 + \sin(2\pi t)] \sum_{j = 1}^{N} \sigma_j^x,
\end{equation}
and is thus rewritten in terms of the bosonic operators as
\begin{equation}
H = \frac{J}{N} \left(n_\uparrow - n_\downarrow\right)^2 - \pi h [1 + \sin(2\pi t)] (a_\uparrow^\dagger a_\downarrow + a_\downarrow^\dagger a_\uparrow).
\label{eq. Htmp}
\end{equation}
We elaborate on the first term of Supplementary Eq.~\eqref{eq. Htmp} using $m = \frac{n_\uparrow - n_\downarrow}{N}$ and noting that
\begin{equation}
\begin{cases}
\frac{n_\uparrow}{N} = \frac{1 + m}{2} \\
\frac{n_\downarrow}{N} = \frac{1 - m}{2}
\end{cases}
\qquad
\Rightarrow
\qquad
\frac{n_\uparrow n_\downarrow}{N^2} = \frac{1-m^2}{4},
\end{equation}
from which
\begin{equation}
m^2 = \left(\frac{n_\uparrow - n_\downarrow}{N} \right)^2 = \frac{n_\uparrow^2 + n_\downarrow^2 - 2n_\uparrow n_\downarrow}{N^2} = \frac{n_\uparrow^2 + n_\downarrow^2}{N^2} +\frac{m^2}{2} - \frac{1}{2},
\end{equation}
and, isolating $m^2$,
\begin{equation}
m^2 = 2\frac{n_\uparrow^2 + n_\downarrow^2}{N^2} - 1 =
2\frac{n_\uparrow(n_\uparrow - 1) + n_\downarrow(n_\downarrow-1)}{N^2} - 1 + \frac{2}{N}.
\end{equation}
Setting $U = 4J$ and $\tau (t) = \pi h [1 + \sin 2\pi t]$, up to irrelevant additional constant terms, the Hamiltonian thus reads
\begin{equation}
H(t) = -\tau(t)(a_\uparrow^\dagger a_\downarrow + a_\downarrow^\dagger a_\uparrow) + \frac{U}{2N}\big(n_\uparrow(n_\uparrow-1) + n_\downarrow(n_\downarrow-1)\big),
\label{eq: DW H}
\end{equation}
where $n_\uparrow = a_\uparrow^\dagger a_\uparrow$ and $n_\downarrow = a_\downarrow^\dagger a_\downarrow$. That is, the LMG model in the symmetric subspace is mapped to a model for $N$ bosons in a double well (the two wells being labeled by $\uparrow$ and $\downarrow$).

\subsection{Equations of motion}
The bosonic representation of Supplementary Eq.~\eqref{eq: DW H} is particularly convenient to obtain dynamical equations. The Heisenberg equations for the bosonic operators read ($\hbar = 1$)
\begin{equation}
\begin{aligned}
\frac{da_\uparrow}{d(it)} &= [H(t), a_\uparrow] = \tau(t) a_\downarrow - \frac{U}{N}n_\uparrow a_\uparrow, \\
\frac{da_\downarrow}{d(it)} &= [H(t), a_\downarrow] = \tau(t) a_\uparrow - \frac{U}{N}n_\downarrow a_\downarrow.
\end{aligned}
\end{equation}
In the limit $N\rightarrow \infty$, upon replacing $a_\uparrow \rightarrow \sqrt{N} \psi_\uparrow$ and $a_\downarrow \rightarrow \sqrt{N} \psi_\downarrow$, with $\psi_\downarrow$ and $\psi_\downarrow$ complex fields, we finally derive the following Gross-Pitaevskii equation
\begin{equation}
\begin{aligned}
\frac{d\psi_\uparrow}{d(it)} &= \tau(t) \psi_\downarrow - U |\psi_\uparrow|^2 \psi_\uparrow,\\
\frac{d\psi_\downarrow}{d(it)} &= \tau(t) \psi_\uparrow - U |\psi_\downarrow|^2 \psi_\downarrow.
\end{aligned}
\label{eq. GPE_SI}
\end{equation}
For an operator $\hat{O} = f(a_\uparrow, a_\downarrow, a_\uparrow^\dagger, a_\downarrow^\dagger)$ written as a function $f$ of the bosonic operators, the beyond-mean-field dynamics of the expectation value $O(t) = \langle \hat{O} \rangle (t)$ can be generally computed within a Truncated Wigner approximation (TWA) as
\begin{equation}
O (t) \approx \langle f(\psi_\uparrow(t), \psi_\downarrow(t), \psi_\uparrow^*(t), \psi_\downarrow^*(t)) \rangle_{\psi_\uparrow(0), \psi_\downarrow(0)}
\label{eq. TWA}
\end{equation}
where $\langle \dots \rangle_{\psi_\downarrow(0), \psi_\uparrow(0)}$ denotes the average over an ensemble of stochastic semiclassical initial conditions $\psi_\uparrow(0)$ and $\psi_\downarrow(0)$ that are drawn according to the quantum initial condition, and then evolve in time with the GPE \eqref{eq. GPE_SI}.

In particular, let us consider as initial condition the symmetrized state $\ket{\psi'(0)}$ with magnetization $m' = 1 - \delta$, with $0 < \delta \ll 1$ and $m'N$ integer (which, since we assume $N\rightarrow\infty$, does not restrict the possible values for $m'$)
\begin{equation}
\ket{\psi'(0)} = \ket{n_\uparrow', n_\downarrow'},
\qquad n_\uparrow' = m'N, \ n_\downarrow' = (1-m')N,
\end{equation}
for which the TWA is performed considering the following ensemble of initial conditions
\begin{equation}
\psi'_\uparrow(0) = \sqrt{1-\frac{\delta}{2}}e^{i\theta_\uparrow(0)},
\qquad
\psi'_\downarrow(0) = \sqrt{\frac{\delta}{2}} e^{i\theta_\downarrow(0)}, 
\label{eq. TWAIC 0}
\end{equation}
with $\theta_\uparrow(0)$ and $\theta_\downarrow(0)$ independent uniform random numbers between $0$ and $2\pi$. Thanks to a gauge transformation, we can always change the initial conditions \eqref{eq. TWAIC 0} into
\begin{equation}
\psi'_\uparrow(0) = \sqrt{1-\frac{\delta}{2}},
\qquad
\psi'_\downarrow(0) = \sqrt{\frac{\delta}{2}} e^{i\theta_0}, 
\label{eq. TWAIC}
\end{equation}
where $\theta_0$ is also a random number between $0$ and $2\pi$. Consider now the limit of $\delta \rightarrow 0$, that is of $\ket{\psi'(0)} \rightarrow \ket{\psi(0)} = \ket{\uparrow, \uparrow, \dots, \uparrow}$. In this limit, the ensemble of stochastic initial conditions \eqref{eq. TWAIC} shrinks in the phase space of complex coordinates $\psi'_\uparrow(0)$ and $\psi'_\downarrow(0)$ towards the point $\psi_\uparrow(0) = 1 - \psi_\downarrow(0) = 1$. If the GPE is nonchaotic, the points of the shrinked ensemble follow close trajectories, so that the TWA average in Supplementary Eq.~\eqref{eq. TWA} can be actually replaced by the evaluation of $f$ for a single trajectory of the ensemble, say the one starting in $\psi_\uparrow(0) = 1 - \psi_\downarrow(0) = 1$. In contrast, if the GPE is chaotic, because of sensitivity to initial conditions, the ensemble quickly spreads, scrambling across the classical phase space. In this case the ensemble trajectories at long-times in \eqref{eq. TWA} interfere destructively, washing out any time oscillation of $O$: the TWA results in thermalization.

In the limit $\ket{\psi'(0)} \rightarrow \ket{\psi(0)} = \ket{\uparrow, \uparrow, \dots, \uparrow}$, it is therefore convenient to consider the following 'single-shot GPE' \cite{pizzi2019period}, to be run just once
\begin{equation}
O(t) \rightarrow \left( f(\psi_\uparrow(t), \psi_\downarrow(t), \psi_\uparrow^*(t), \psi_\downarrow^*(t)) \right)_{\psi_\uparrow(0) = 1, \psi_\downarrow(0) = 0},
\label{eq. SSGPE}
\end{equation}
which is then expected to be accurate when nonchaotic, and to signal quantum thermalization when chaotic, which motivates the use of the symbol $``\rightarrow"$, rather than $``="$.

In particular, considering the observable $\vec{S} = \frac{1}{N} \sum_{j = 1}^{N} \vec{\sigma}_j$, from Supplementary Eq.~\eqref{eq. map} we thus write
\begin{equation}
\frac{\langle S^x \rangle + i \langle S^y \rangle}{2} \rightarrow \psi_\uparrow^* \psi_\downarrow,
\qquad
\langle S^x\rangle \rightarrow |\psi_\uparrow|^2 - |\psi_\downarrow|^2.
\end{equation}

\section*{Supplementary Note 2: EXACT DIAGONALIZATION}
\begin{figure}[t]
	\begin{center}
		\includegraphics[width=\linewidth]{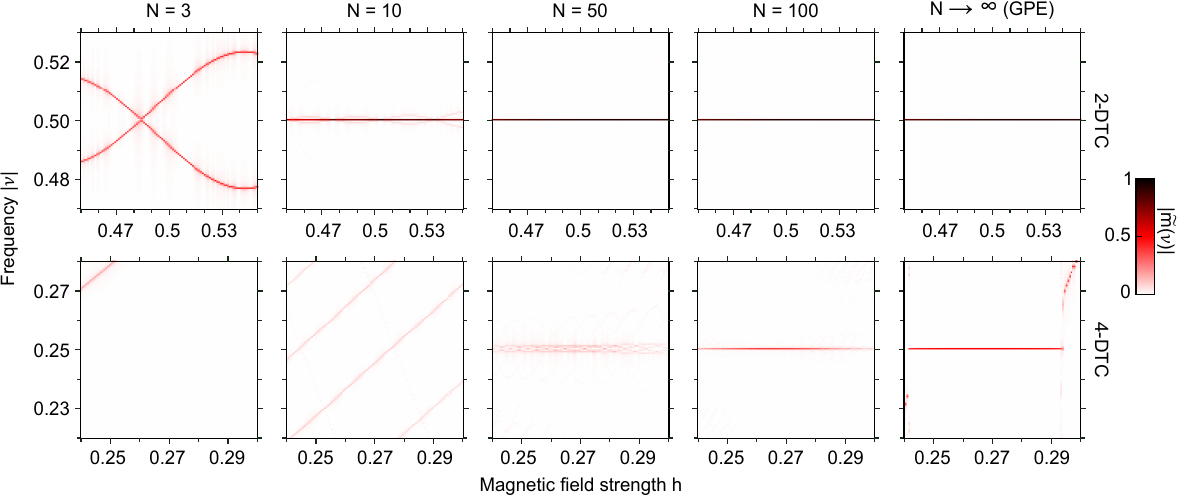}\\
	\end{center}
	\vskip -0.5cm \protect\caption
	{\textbf{Exact diagonalization and system size scaling}. For an initially $\textbf{z}$-polarized state $\ket{\psi(0)} = \ket{\uparrow, \uparrow, \dots, \uparrow}$ we plot the Fourier transform $\tilde{m}(\nu)$ of the magnetization $m = \langle \sigma_j^z \rangle$ for various number of spins $N$ and a fixed interaction strength $J = 0.5$. Results from the GPE in the limit $N \rightarrow \infty$ are reported on the right as a reference. (top row) For $h \approx 0.5$, the $2$-DTC plateau at frequency $0.5$ emerges for increasing $N$, being clearly visible already for $N \gtrapprox 10$. (bottom row) For $h \approx 0.25$, the $4$-DTC plateau at frequency $0.25$ also emerges for increasing $N$, but it does so only for considerably larger system sizes $N \gtrapprox 100$, which might explain the difficulties in observing higher-order DTCs.}
	\label{figS1}
	\vskip -0.5cm
\end{figure}

The LMG model is also particularly suitable for exact diagonalization techniques. In fact, the size of the symmetric subspace ($N+1$) scales only linearly with the system size $N$, making exact diagonalization viable for systems which are much larger than the ones typically considered in less-symmetric 1D models. In Fig.~\ref{figS1} we show the Fourier transform $\tilde{m} (\nu)$ of the magnetization $m = \langle \sigma_j^z \rangle$ for an initially $\textbf{z}$-polarized state $\ket{\psi(0)} = \ket{N_\uparrow} = \ket{\uparrow, \uparrow, \dots, \uparrow}$. For increasing $N$, we can observe the emergence of the constant-frequency plateaus signaling the $n$-DTCs. First, this confirms that the GPE captures the correct dynamics for large $N$, as expected. Second, we notice that, for the $2$-DTC and the $4$-DTC, the plateau is clearly visible only for $N \gtrapprox 10$ and $N \gtrapprox 100$, respectively. This is remarkable, as it suggests that higher-order DTCs may be numerically harder to observe as they could appear at larger system sizes, generally beyond the reach of exact diagonalization techniques. This observation might explain why higher-order DTCs have so far remained elusive to numerical examples \cite{machado2019prethermal}.

\section*{Supplementary Note 3: SPIN-WAVE APPROXIMATION}
In this Note, we briefly summarize the idea behind the spin-wave approximation, which is thoroughly explained in Refs.~\cite{lerose2018chaotic, lerose2019impact} and the supplementary material therein, to which we redirect the reader for further details.

In a DTC evolving from an initially $\textbf{z}$-polarized state $\ket{\psi(0)} = \ket{\uparrow, \uparrow, \dots, \uparrow}$, the spins are mostly aligned at all times. Imperfections in the alignment can be described within a Holstein-Primakoff transformation in terms of bosonic spin-wave quasiparticles $b_k^\dagger$. Crucially, the collective spin $\vec{S} = \frac{1}{N} \sum_{j=1}^{N} \vec{\sigma}_j$ rotates as a function of time in the lab frame. Therefore, the Holstein-Primakoff transformation has to be performed in a rotating frame $(X, Y, Z)$ such that the, say, $Z$ axis tracks the orientation of the collective spin at all times. This tracking is encoded in the condition $\langle S^X \rangle = \langle S^Y \rangle = 0$, from which the dynamics of the rotating frame is obtained self-consistently. On top of this, an approximation is made in that the Hamiltonian is expanded to lowest non-trivial order in the density of spin-wave excitations $\epsilon = \frac{2}{N} \sum_{k \neq 0}^{N} \langle b_k^\dagger b_k \rangle$, which should remain $\ll 1$ for the approximation to be consistent. This procedure results in a set of $2N$ ordinary differential equations similar to Eqs.~(26) and (29) in the Supplemental Material of Ref.~\cite{lerose2018chaotic}, describing the rotation of the new reference frame and the dynamics of the spin waves at the various momenta.

Finally, we remark the main differences between the implementation of the spin-wave approximation in our work and in Ref.~\cite{lerose2018chaotic}. First, Ref.~\cite{lerose2018chaotic} considers a constant-in-time Hamiltonian, whereas the parameters of the Hamiltonian in the present work are time-dependent. As a consequence, the parameters in the system of ordinary differential equations become time-dependent. Second, Ref.~\cite{lerose2018chaotic} considers a nearest-neighbor interaction on top of a fully-connected one, whereas we consider the more general case of nearest-neighbor interaction on top of a power-law one. The $\cos k$ that appears in the equations of Ref.~\cite{lerose2018chaotic} is therefore substituted by a more generic $\tilde{\mathcal{J}}_k = \sum_{j = 1}^{N} \mathcal{J}_{r_{1,j}} e^{-i r_{1,j} k}$ in ours, where $\mathcal{J}_{r_{i,j}}$ contains both the nearest-neighbor and the power-law interactions.

\section*{Supplementary Note 4: BINARY DRIVING}
\begin{figure}[t]
	\begin{center}
		\includegraphics[width=0.5\linewidth]{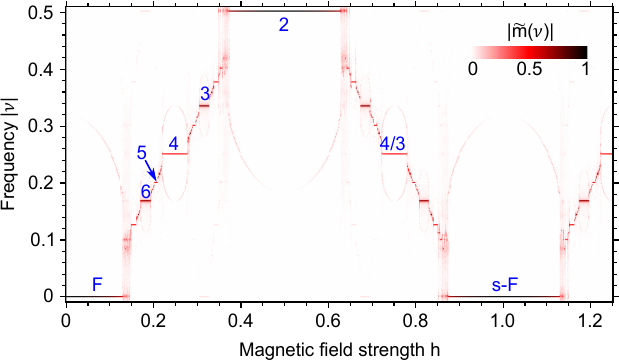}\\
	\end{center}
	\vskip -0.5cm \protect\caption
	{\textbf{Phase diagram fragmentation for a binary Floquet driving}. This figure is in complete analogy with Fig.~1b of the main paper, but considers the binary Floquet protocol \eqref{eq. binary}. We plot the Fourier transform $\tilde{m}(\nu)$ of the magnetization $m = \langle \sigma_j^z \rangle$ in the plane of the transverse field strength $h$ and of the frequency $\nu$, for a fixed interaction strength $J = 0.5$ and computed over $500$ periods. We observe that the spectral lines fragment in plateaus with constant frequency, signaling the $n$-DTCs, (dynamic) ferromagnet and stroboscopic-ferromagnet. In blue, we indicate the index $n$ of some of the resolved DTCs.}
	\label{figS2}
	\vskip -0.5cm
\end{figure}

In this Note, we compliment the results from the main paper showing that the fragmentation of the phase diagram to host a multitude of higher-order DTCs also occur for a binary Floquet protocol. In particular, we consider a periodic Hamiltonian with period $1$ alternating fully-connected interaction and transverse field
\begin{equation}
H(t) =
\begin{dcases}
+ 2 \frac{J}{N} \sum_{i,j = 1}^{N} \sigma_i^z\sigma_j^z
\quad &0 < t \le 0.5 \\
- 2 \pi h \sum_{j = 1}^{N} \sigma_j^x
\quad &0.5 < t \le 1.
\end{dcases}
\label{eq. binary}
\end{equation}
From the Hamiltonian \eqref{eq. binary}, in the limit $N \rightarrow \infty$, we can derive a GPE in complete analogy with Supplementary Eq.~\eqref{eq. GPE_SI}. Solving it for the initially $\textbf{z}$-polarized state $\ket{\psi(0)} = \ket{\uparrow, \uparrow, \dots, \uparrow}$, we obtain the spectrum $\tilde{m}(\nu)$ of Fig.~\ref{figS2}. Also for this model, it is possible to check within a spin-wave approximation that the dynamical phases persist when deviating from the LMG limit of fully-connected spins, as long as the interaction range is large enough, in complete analogy with the results of Fig.~4 of the main paper.

\section*{Supplementary Note 5: STABILITY AGAINST $\mathbb{Z}_2$ SYMMETRY-BREAKING TERMS}
The Hamiltonian $H(t)$ of Eq.~1 in the main text is $\mathbb{Z}_2$ symmetric, since its commutator with a global spin flip $\prod_{j = 1}^{N} \sigma_j^x$ vanishes. For a $2$-DTC, a question raises whether the subharmonic response truly stems from the breaking of the time-translational symmetry, or it rather ``piggybacks'' on an underlying breaking of the $\mathbb{Z}_2$ symmetry\cite{von2016absolute, else2017prethermal, else2020long}. This question motivates an analysis to assess whether the $2$-DTC is stable in the presence of a term breaking the underlying $\mathbb{Z}_2$ symmetry explicitly, such as a longitudinal field. For higher-order and fractional $n$-DTCs, instead, no such an issue exist, because the Hamiltonian lacks any $\mathbb{Z}_n$ symmetry $(n > 2)$, and the DTCs must be a ``genuine'' manifestation of time-symmetry breaking alone, breaking an \emph{emergent} $\mathbb{Z}_n$ symmetry. Nonetheless, it is still interesting to investigate the effects that the explicit breaking of the $\mathbb{Z}_2$ symmetry has on the $n$-DTCs. For completeness, in this Note, we therefore perform a stability analysis upon introducing a longitudinal field.

Consider the following Hamiltonian in which, on top of the Hamiltonian $H(t)$ from Eq.~(1) in the main text, we add a longitudinal field of strength $\pi h_z$ (the factor $\pi$ comes from the analogy with the definitions for the transverse field)
\begin{equation}
H'(t) = H(t) + \pi h_z \sum_{j = 1}^{N} \sigma_j^z
\label{eq. long field}
\end{equation}

\begin{figure}[t]
	\begin{center}
		\includegraphics[width=\linewidth]{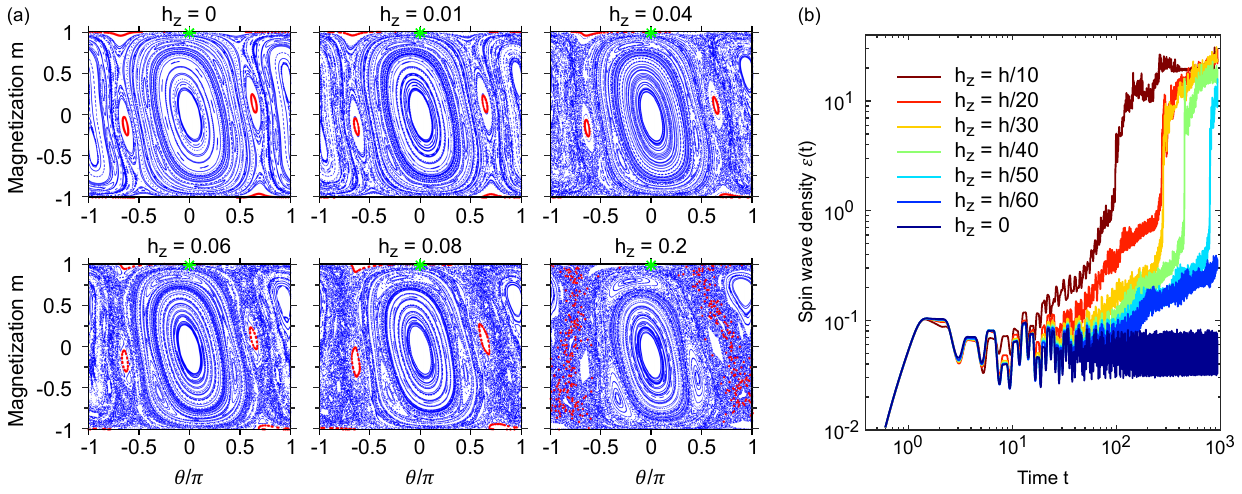}\\
	\end{center}
	\vskip -0.5cm \protect\caption
	{\textbf{Stability against $\mathbb{Z}_2$ symmetry-breaking terms}. We investigate the stability of the $4$-DTC in the presence of a longitudinal field of strength $h_z$ in Supplementary Eq.~\eqref{eq. long field}. (a) In the LMG limit, the dynamics is analysed with Poincaré maps analogue to those of Fig.~2 in our manuscript. If $h_z$ is small enough, the islands underpinning the $4$-DTC qualitatively do not change, and the $4$-DTC is stable. For increasing $h_z$, larger and larger regions of the phase space become chaotic, and the $4$-DTC is eventually broken. (b) Dynamics of the spin-wave density $\epsilon(t)$ for $\alpha = 1.2$, $\lambda = 0.03$, and $N=1000$. The prethermal nature of the DTC is manifest in this case: spin waves proliferate after a time roughly scaling as $\sim e^{1/h_z}$. In both (a) and (b), we considered $h=0.26$ and $J=0.5$.}
	\label{figS3}
	\vskip -0.5cm
\end{figure}

In Fig.~\ref{figS3}, we investigate the stability of the $4$-DTC in the presence of the longitudinal field $h_z$ breaking the $\mathbb{Z}_2$ symmetry of the Hamiltonian. In the LMG limit ($\alpha = \lambda = 0$), the introduction of a small field $h_z$ has qualitatively no significant effect. In the Poincaré map, the $4$ islands characterizing the $4$-DTC are in fact just slightly deformed for a small $h_z$, suggesting that the subharmonic response lasts up to infinite time, similarly to $h_z=0$. Away from the LMG limit, instead, the dynamics within the spin-wave approximation shows a proliferation of the spin waves after a long prethermal regime. Although it is hard to assess the accuracy of this approximate method, the results suggest that the expected prethermal nature of the DTC beyond the LMG limit is only changed quantitatively, with the heating time scaling exponentially as $t_{th} \sim e^{1/h_z}$. For $h_z=0$ we recover the results for which the 4-DTC appears infinitively long-lived within the spin-wave approximation.

\end{document}